%% file: main.tex
\documentclass[twocolumn]{aastex62}
\usepackage{lipsum}
\usepackage{graphicx}
\usepackage{xfrac}
\usepackage{amssymb}
\usepackage{xcolor}
\usepackage{ tipa }
\usepackage{tablefootnote}
\usepackage{makecell}
\usepackage{url}

\input{shorthand.tex}

\shorttitle{NSC Solar System I}

\begin{document}

\title{\editoneb{Exploring the Solar System with the NOIRLab Source Catalog I:} Detecting \eonesout{Solar System }Objects \eonesout{in NSC DR1 }with CANFind}
\input{authors.tex}

\begin{abstract}

Despite extensive searches and the relative proximity of solar system objects (SSOs) to Earth, many remain undiscovered and there is still much to learn about their properties and interactions. \etwosout{In }This work \edittwob{is the first in a series dedicated to detecting and analyzing SSOs in the all-sky NOIRLab Source Catalog (NSC).  We }search the first data release of the \etwosout{NOIRLab Source Catalog (}NSC\etwosout{)} \etwosout{for SSOs using}\edittwob{with} CANFind, a Computationally Automated NSC tracklet Finder.  NSC DR1 contains 34 billion measurements of 2.9 billion unique objects, which CANFind categorizes as belonging to ``stationary'' (distant stars, galaxies) or moving (SSOs) objects via an iterative clustering method. Detections of stationary bodies for proper motion $\mu\leq$2.5\arh \editoneb{(0.017\ded)} are identified \eonesout{to be}\editoneb{and} analyzed separately.  \etwosout{and r}\editoneb{R}emaining detections belonging to high-$\mu$ objects are clustered together over single nights to form\etwosout{ structures called} ``tracklets''. Each tracklet contains detections of an individual moving object, and is validated based on spatial linearity and motion through time. Proper motions are then calculated and used to connect tracklets and other unassociated measurements over multiple nights by predicting their locations at common times, forming ``tracks''.  This method extracted \eonesout{524,501}\editoneb{527,055} tracklets from NSC DR1 in an area covering \eonesout{27,882}\editoneb{29,971} square degrees of the sky.  The data show distinct groups of objects with similar \editoneb{observed} $\mu$ in ecliptic coordinates, namely Main Belt Asteroids, Jupiter Trojans, and Kuiper Belt Objects.  Apparent magnitudes range from 10--25 mag in the \textit{ugrizY} and \textit{VR} bands.  Color-color diagrams show a bimodality of tracklets between primarily carbonaceous and siliceous groups, supporting prior studies.  

\end{abstract}


\section{Introduction} \label{sec:intro}

Although the sky has been extensively searched for solar system objects (SSOs) over the last few hundred years, new bodies are still being found today.  \eonesout{From}\editoneb{Long before} the first discovery of an asteroid (Ceres) at the Palermo Observatory in 1801 \citep{cunningham09} until as recently as 1991, the process of identifying \eonesout{Solar System Objects} \eonesout{(}SSOs\eonesout{)} depended entirely on the human eye.  Through the origin of astronomical surveys specifically intended for SSO detection, and later the addition of charge-coupled devices (CCDs) for astronomy purposes by the Spacewatch Telescope at the Steward Observatory \citep{gehrels91}, SSOs were spotted either by direct observation or by ``blinking'' between physical exposures of the sky in a lengthy process. This changed when \cite{Rabinowitz91} created the first automated method used for detecting Earth-approaching asteroids in near real time using the Spacewatch Telescope.  With sequential exposures taken by the telescope, the detection software identified trails that moving objects (MOs) left in the first two exposures, and displacements between them and a third exposure.  This new method of MO detection marked the beginning of a striking growth in the rate of asteroid discovery, and of our knowledge of the solar system.

SSOs fall into categories based on their orbital parameters and composition, called ``groups'' and ``families''.  While groups have similar orbits, families are more closely related as they often consist of fragments from past collisions. \eonesout{One of the broader groups is the}\editoneb{Distinct to us by their high proper motion and perihelia $\leq$ 1.3 \textit{AU}, }Near Earth Objects (NEOs)\eonesout{, which have perihelia of less than 1.3 \textit{AU} and} are potential impactors due to their proximity.  It wasn't until 2008 that the first detection of an asteroid before its impact was made by the Catalina Sky Survey \citep[CSS,][]{Kowalski08}.  Various organizations and warning systems like the Asteroid Terrestrial-impact Last Alert System \citep[ATLAS,][]{atlas} are dedicated to the NEO search, as even a relatively small asteroid has the potential for mass destruction \citep{dearborn20} and lasting effects on the Earth’s environment \citep{marchi19,koeberl19}.  To address the problem, in 1991 the Spaceguard Survey Network was established by NASA in response to direction from the United Stated Congress to increase the detection rate of asteroids which cross Earth's orbit \citep{Morrison}.  Spaceguard has led to a significant increase in surveys dedicated to discovering NEOs, and in 2011 the NEO subset of the Wide-Field Infrared Survey Explorer \citep[NEOWISE,][]{Mainzer} reported the completion of the initiative to identify 90\% of NEOs greater than 1 km in diameter.  The search expanded in 2005 when NASA was instructed to find 90\% of NEOs with diameters greater than 140 m within 15 years, a time window which is closing.  Despite CSS, NEOWISE, ATLAS, \edittwob{the Pan-STARRS Survey \citep{Denneau13}} and further programs and surveys dedicated to reaching this goal, surprise impacts continue; the 2013 event in Chelyabinsk was caused by an undetected NEO with an estimated diameter of 20 m \citep{popova13}.  \editoneb{This goes to show that blind spots still exist in our tracking methods and additional SSO exploration is, at the very least, beneficial.}

Observational limitations also make it particularly difficult to detect many of the smaller subsets within the NEO population, including Atens \editoneb{(Earth-crossing orbit)}, Atiras \editoneb{(orbits within Earth's)}, Vatiras\editoneb{ (orbits within Venus's)}, and Earth Trojans \editoneb{(co-orbiting with Earth)}.  Due to their close proximity to the sun, members of these groups elude observation and it is estimated that only 5--7\% of Atiras with absolute magnitudes less than 20\eonesout{th mag} \editoneb{($d\approx$ 300 $m$)} have been found \citep{ye20}.  The least studied groups are the Vatiras and the Earth trojans, as only one of each has been discovered as of 2020 \citep{Greenstreet20,Connors11}. \editoneb{Studies such as} \cite{Markwardt}\footnote{Data from \cite{Markwardt} are included in the NSC and were analyzed as a part of this study.},\editoneb{ which }searched for Earth trojans using exposures from DECam,\editoneb{ sometimes require confining the area included in their search in order} \eonesout{confining the survey to 8 fields covering an area of 24 square degrees }to optimize the chances of detection.  This and other surveys that constrain their fields to small sections of the sky \editoneb{to improve detection prospects}, like those performed by \cite{Terai13} and \cite{ivezic01}, \editoneb{often} do not allow for orbit determination as the observation arcs \editoneb{observed were}\eonesout{are} too short to extrapolate. 

Further out beyond the orbit of Mars, the Main Belt is filled with dozens of dynamic asteroid groups and families that can be constrained by searches through large catalogs \citep{lemaitre}.  Main Belt Asteroids (MBAs) observed by the Sloan Digital Sky Survey are shown to have a bimodality of carbonaceous and rocky siliceous objects \citep{ivezic01}, with size distributions estimated assuming consistent albedos throughout each population.  \editoneb{MBA size and color can be relatively well-studied because of their abundance and our closeness, but contrasting results raise questions.}  The steepest \editoneb{estimated size distribution} is a simple power law $\sim$ 0.9 from the 2015 High cadence Transient Survey (HiTS) campaign \citep{pena}, similar only to the results of \cite{parker}, who also use SDSS data.  However, the 2014 HiTS campaign reported a slope of $\sim$ 0.76 in the \textit{g} band for brighter objects and $\sim$ 0.28 for faint objects \citep{pena}, which is closer to more common results like those found by \cite{gladman09}.  According to \cite{pena}, this is likely due to an interesting result described by \cite{bhattacharya}, who reported that objects at ecliptic latitudes \editoneb{\textit{b}} $\geq$ 5$^{\circ}$ observed by the Spitzer Space Telescope were 30\% fainter than SSOs in the ecliptic plane (the 2015 HiTS campaign observed fields at a higher ecliptic latitude than the 2014 campaign).  \editoneb{Additional SSOs must be observed to model a more accurate size distribution, and to provide more insight on its relation to ecliptic latitude.}

At the far reaches of our Solar System, Trans-Neptunian Objects (TNOs) are \eonesout{still gravitationally bound to the Sun but far enough away from it}\editoneb{distant enough} (semimajor axis $a\gtrsim$ 30 \textit{AU}) that they do not feel the effects of the main planets as much as objects in the inner Solar System.  This means that they are particularly susceptible to external forces like galactic tides, passing stars, or perhaps even a distant unknown planet.  These interactions with extrasolar forces provide insight into the dynamics, formation, and history of the Solar System.  TNOs, which include Kuiper Belt Objects (KBOs) and Oort Cloud Objects (OCOs), are essentially remnants of the primordial Solar System, and studying their properties will provide stronger constraints on planet formation theories.

The potential presence of an unseen planet or mass has been explored by several authors after the discovery of Neptune and its perturbed orbit.  Pluto's discovery in 1930 confirmed the existence of TNOs and alluded to the possibility of another larger object, dubbed ``Planet X'' by astronomer Percival Lowell.  The most recent and promising hypothesis was established when \cite{Brown04} discovered 90377Sedna (Sedna), the most distant SSO known at the time.  Sedna has an unusually eccentric and inclined orbit, and one of the proposed causes of this is the gravitational scattering by an unseen planet.  
This hypothesis was further developed when \cite{Trujillo14} discovered the second Sedna-like body, 2012 VP$_{133}$, orbiting at the very edge of the Kuiper belt using the Dark Energy Camera (DECam, \citealt{flaugher}).  They found that both Sedna and 2012 VP$_{133}$ showed clustering in the argument of perihelion ($\omega$), the angle between an inclined SSO's perihelion and the point where its orbital plane intersects with the ecliptic plane.  The mechanism by which this clustering is thought to be achieved is referred to as ``gravitational shepherding'', which describes the effect of a large body on the orbits of nearby smaller bodies.  As a small body passes by a larger one, the latter’s nonzero net gravitational force on the former alters its trajectory and shifts its orbit.  To this day, the gravitational shepherding due to a distant massive planet, often referred to as ``Planet 9'', remains by far the best explanation for the clustering seen in extra-TNOs.

\cite{batygin16} added to this hypothesis by showing a clustering in not only $\omega$ of Kuiper Belt orbits, but a physical clustering of their perihelia as well.  They also report that the orbits of known KBOs with semimajor axis $a\geq227\,AU$  all approach perihelion within 94$^{\circ}$ of longitude of each other, and all lie in an orbital plane $\sim22^{\circ}$ from the ecliptic \citep{brown16}.  These numbers are supported by the recent discovery of 541132 Lele\={a}k\={u}honua (2015 TG387) by \cite{shep19}.  Two distinct groups of TNOs with similar $\omega$ have allowed astronomers to derive theoretical orbits of Planet 9.

Alternative explanations have been proposed and rejected; \cite{khain18} showed that the clustering may be due in part to the primordial state of the Solar System, but would not have produced an effect as substantial as the one observed.  \cite{brown17} effectively ruled out observational bias as the cause, estimating that the probability that the 10 KBOs known at the time with $a\geq$ 230 \textit{AU} were part of a larger population with uniform $\omega$ was 1.2\%.  A later work \citep{brown19} reported a 0.02\% probability that distant KBOs would be clustered as strongly as observed in both $\omega$ and in orbital pole position if the sole reason for this clustering was a detection bias. 

Constraints on a mass and orbit for Planet 9 have changed over the years, from \cite{Hogg91} placing a detectable limit on its mass of about 70 M$_{\oplus}$, to \cite{batygin19} more recently constricting it to $\sim$5-10 M$_{\oplus}$ thanks to the greater number of known TNOs.  Through a numerical simulation, \cite{brown16} suggested that the perihelion of Planet 9 would be between 150 and 350 \textit{AU}, with an inclination within 30$^{\circ}$ of the ecliptic.  

Studies of the Kuiper Belt and searches for TNOs  have yielded small bodies and dwarf planets, but no body large enough to cause the observed clustering of TNOs has been found. Although the existence of Planet 9 itself remains unconfirmed, the discovery of additional TNOs through the work described in this paper could provide further evidence.  Estimations of its parameters require a working knowledge of the most distant SSOs, hence the need to continue the search for them.  Similarities between new and known objects may support the evidence of Planet 9 and contribute to estimations of its properties, while differences may lead to new hypotheses for unidentified forces acting on TNOs.  


Two common obstacles that surveys face include limitations on the brightness of what can be detected (typical depth $\approx$ 19--24 mag, \citealt{Denneau13}) and the area of the sky covered by observations.  SSO-based surveys are \eonesout{typically}\editoneb{often} constrained to the ecliptic where the bulk of the solar system's mass is found, meaning more surveys and catalogs with more sky coverage must be searched to detect objects with high orbital inclinations to explore relationships such as the latitude--magnitude dependence shown by \cite{pena} and \cite{bhattacharya}.  Several wide-field surveys have been used to achieve this, such as CSS, the Near-Earth Asteroid Tracking program \citep[NEAT;][]{Pravdo99}, CCS, and the the Lincoln near-Earth asteroid program \citep[LINEAR; ][]{stokes00}, although both NEAT and LINEAR focus mainly on the northern celestial hemisphere due to the location of the telescopes used.  The upcoming Legacy Survey of Space and Time \citep[LSST;][]{ivezic19} with the Rubin Observatory will extensively monitor the southern sky and is predicted to increase the known number of all small body populations by 10--100$\times$ \citep{jones16}, and is expected to be a significant detector of NEOs \cite{Veres17}.  However, much of the northern sky will remain unmonitored at the faint magnitudes.  

Independent endeavors to create detection algorithms that can be applied to any astronomical data, for example the NEARBY Platform \citep{stefanut19}, the Moving Object Processing System (MOPS) developed for Pan-STARRS, and other methods (e.g. \citealt{lieu19}, \citealt{perdelwitz18}, and \citealt{holman18}) have varying degrees of success and can typically be tuned to find specific families of SSOs with similar orbits.  A recent example of this is THOR (Tracklet-less Heliocentric Orbit Recovery), an algorithm developed with detections from the Zwicky Transient Facility to extract SSO detections by applying test orbits to the data \citep{Moeyens}.  

\begin{figure*}[ht] 
    \centering
    \includegraphics[scale=0.32]{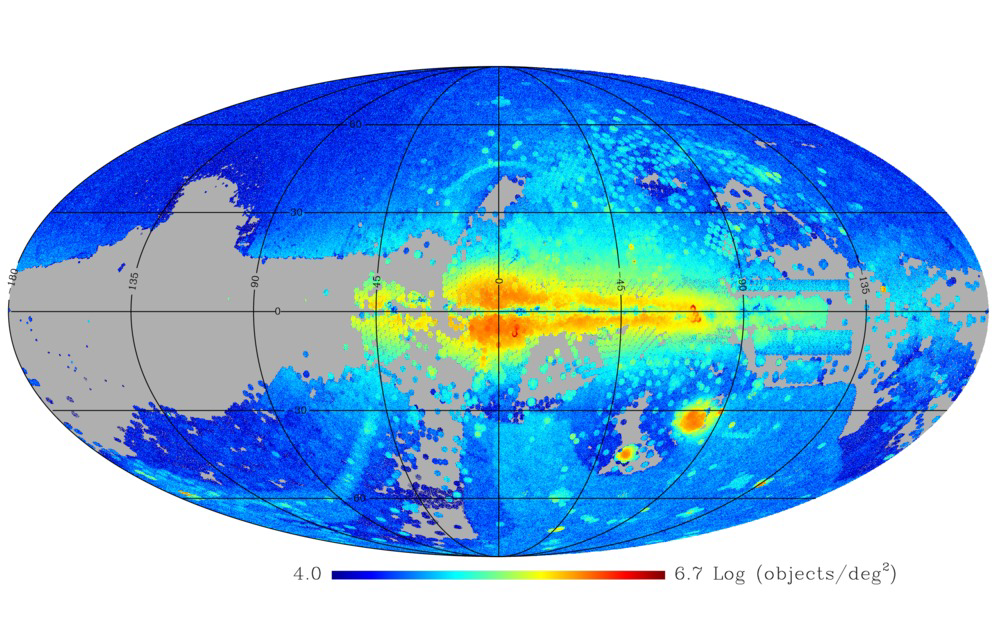}
    \includegraphics[trim={0cm 4cm 2cm 3cm},clip,width=0.47\hsize,angle=0]{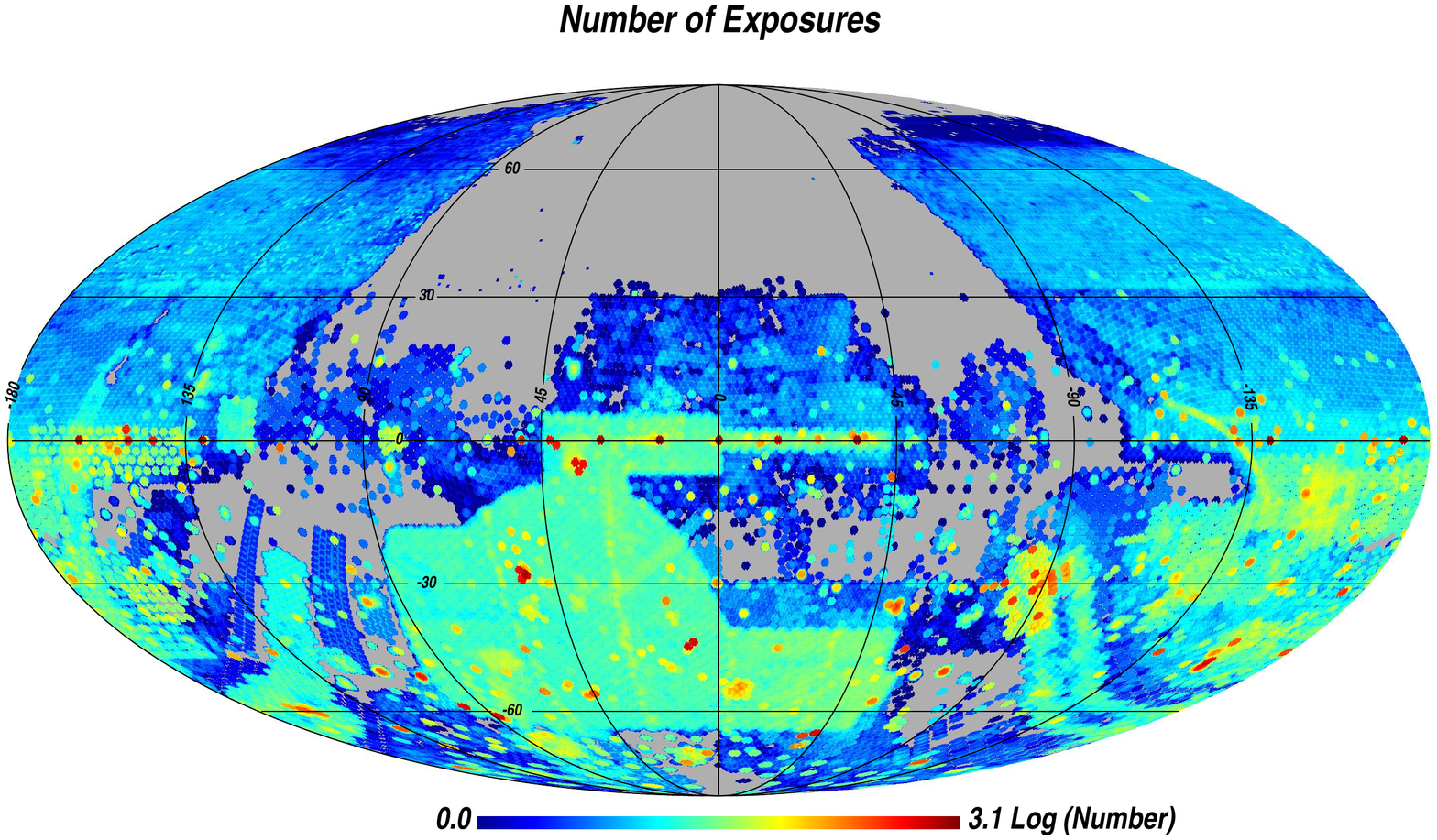}
    \caption{\textit{(Left)} Density plot (number of objects per square degree) of the objects in NSC DR1 \citep{Nid2018} in galactic coordinates.  3/4 of the sky is covered, primarily the southern celestial hemisphere. \textit{(Right)} Number of exposures overlapping any piece of sky in NSC DR1, on a logarithmic scale and in equatorial coordinates.  Most areas have tens of exposures while some have thousands.}
    \label{fig:nsc_data}
\end{figure*}

This work describes the background and implementation of the Computationally Automated NSC-tracklet Finder (CANFind), an approach designed to identify SSOs in photometric catalogs with individual-epoch measurements like the NOIRLab Source Catalog (NSC)\footnote{Formerly known as the NOAO Source Catalog.}.  The NSC is a crowd-sourced catalog covering 3/4 of the sky, with 34 billion observations, most regions are covered by at least 3 exposures.  We test CANFind on the first data release of the NSC, and detect over 500,000 tracklets to magnitudes of 24 mag.  50,222 of these 500,000 tracklets have \textit{g-r}, \textit{r-i} color-color information showing a distinct bimodality of two general asteroid types (carbonaceous and silicious). 85.9\% of tracklets were found within $\pm30^{\circ}$ of the ecliptic plane, and \edittwob{60\% have more than three measurements.}\eonesout{71.6\% had observed proper motions between 25--45\arh \editoneb{(0.17--0.3\ded)}.}  

The paper is organized as follows:  Section \ref{sec:data} describes the first data release of the NOIRLab Source Catalog and its benefits for the use of SSO detection.  CANFind is explained in Section \ref{sec:meth}, an overview of the detection efficiency simulation is in Section \ref{sec:sim}, the results are discussed in Section \ref{sec:results},  and future work to be done with NSC tracklets is described in Section \ref{sec:future}.  \edittwob{This introduces a set of planned publications all involving SSO detection in the NSC and subsequent investigation.  Upcoming projects will connect tracklets with more spatially disperse measurements, fit NSC tracklets to SSO orbits, and analyze recovered SSO populations, morphology, comet activity, and streaked detections.}


\section{NOIRLab Source Catalog}
\label{sec:data}

This project uses the first data release of the NOIRLab Source Catalog \citep[NSC;][]{Nid2018} to detect SSOs. The NSC is a crowd-sourced catalog using public data from the NOIRLab Science Archive processed in a uniform manner. The NSC has 34 billion individual observations of 2.9 billion distinct objects \eonesout{and}\editoneb{from} 255,000 images \editoneb{spanning five years of observation} \edittwob{from 2012--2017}.  Data come from DECam on CTIO-4m, KPNO-4m+Mosaic-3, and Bok-2.3m+90Prime from the DESI Legacy Survey \citep{Dey2019}.  \editoneb{Table \ref{tab:instruments} lists several specifications from the instruments used by the surveys included in the catalog.}  The NSC contains a large amount of time series information that is useful for searches of moving objects and photometrically variable objects.  \editoneb{Table \ref{tab:nscdr1} lists information from the 11 top-contributing surveys in DR1.}

\begin{deluxetable*}{lcccccc}
\caption{\editoneb{Instrument Specifications and coverage in NSC DR1}}
\label{tab:instruments}
\tabletypesize{\footnotesize}
\tablehead{
\colhead{Instrument} & \colhead{Observatory} & \colhead{Coverage(MMYYYY)} & \colhead{$N_{exposures}$} & \colhead{FOV ($^{\circ}$)} & \colhead{CCDs} & \colhead{pixel spacing (\arcsec)}
}
\startdata
DECam\tablenotemark{a} & CTIO Blanco 4m & 092012--082017 & 195489 & 3 & 62 & 0.26 \\
Mosaic-3\tablenotemark{b} & KPNO Mayall 4m & 022016--082017 & 40224 & 3 & 4 & 0.26\\
90Prime\tablenotemark{c} & Steward Bok 2.3m & 022016--062017 & 19741 & 1 & 4 & 0.45 \\
\enddata
\tablenotetext{a}{Dark Energy Camera \cite{Dey2019}; \url{http://www.ctio.noao.edu/noao/node/1033}} 
\tablenotetext{b}{Mosaic-3 Wide Field Imager \cite{mosaic3}; \url{https://www.noao.edu/kpno/mosaic/}} 
\tablenotetext{c}{Prime Focus CCD Mosiac Camera \cite{zou17}}
\end{deluxetable*}

\begin{figure}
\centering
\includegraphics[width=1\hsize]{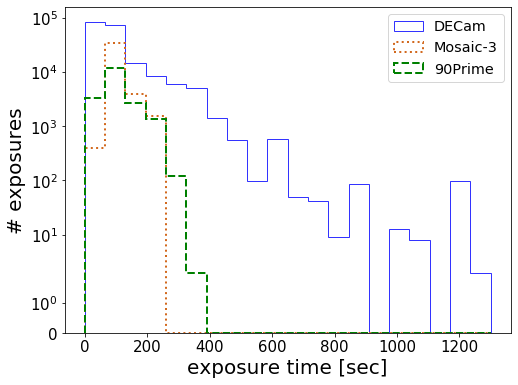}
\caption{\edittwob{Histogram of exposure times for all exposures in the NSC.}  }
\label{fig:exp_hist}
\end{figure}

\begin{deluxetable*}{llp{20mm}c>{\em}cp{25mm}p{20mm}}
\caption{\editoneb{General information for the 11 top contributing surveys in NSC DR1, expanded from \cite{Nid2018} Table 1}.}
\label{tab:nscdr1}
\tablecolumns{7}
\tabletypesize{\footnotesize}
\tablehead{
\colhead{Survey Name} & \colhead{Instrument}& \colhead{PI} & \colhead{$N_{exposures}$} & \colhead{Bands} & \colhead {Depth \textit{(mag)}} & \colhead{PSF FWHM (\arcsec)}
}
\startdata
DES\tablenotemark{a}                & Blanco/DECam    & Friedman                 & 58546 & grizY & 24.5, 24.3, 23.5,\newline 22.9, 21.7  & 1.2, 1.1 ,1.0,\newline 0.9, 0.9\\	
Legacy Surveys:\tablenotemark{b}    & \hspace{1cm}    & \hspace{1cm}             & \hspace{1cm} & \hspace{1cm} & \hspace{1cm} & \hspace{1cm}\\
\hspace{1cm}DECaLS\tablenotemark{c} & Blanco/DECam    & Schlegel\newline \& Dey  & 17533 & grz &24.0, 23.5, 22.5 & 1.3, 1.2, 1.1 \\
\hspace{1cm}BASS\tablenotemark{d}   & Bok/90Prime     & Zou\newline \& Fan       &  19741 & gr &23.7, 23.1 & 1.6, 1.45\\
\hspace{1cm}MzLS\tablenotemark{e}   & Mayall/Mosaic-3 & Dey	                     & 40224 & z & 22.6 & 1.01\\
NEO\tablenotemark{f}                & Blanco/DECam    & Allen                    & 11800 & VR & 23.0 & $\sim$1\\
DECaPS\tablenotemark{g}             & Blanco/DECam    & Finkbeiner               & 7590 & grizY & 23.7, 22.8, 22.2,\newline 21.8, 21.0 & $\sim$1\\
Bulge Surveys:                      & \hspace{1cm}    & \hspace{1cm}             & \hspace{1cm} & \hspace{1cm} & \hspace{1cm} & \hspace{1cm}\\
\hspace{1cm}BDBS\tablenotemark{h}   & Blanco/DECam    & Rich                     &  3849	& ugrizY & 23.5, 23.8, 23.5,\newline 23.1, 22.5, 21.8 & $\sim$0.8--1.8 \\
\hspace{1cm}DSSGB\tablenotemark{i}  & \nodata         & Saha                     &  3837 & ugriz & \nodata & \nodata\\
Light Echoes\tablenotemark{j}       & Blanco/DECam    & Rest                     & 7622 & rVR & \nodata & \nodata\\
SMASH\tablenotemark{k}	            & Blanco/DECam    & Nidever                  & 6645 & ugriz & 23.9, 24.8, 24.5,\newline 24.2, 23.5 & 1.2, 1.1, 1.0,\newline 1.0, 0.9 \\
BLISS\tablenotemark{l}              & Blanco/DECam    & Soares-\newline Santos   & 3049 & griz & \nodata & \nodata\\
\hline\\
NSC DR1                             & \nodata         & Nidever                  & 255,454 & ugrizYVR & 23.1, 23.3, 23.2,\newline 22.9, 22.2, 21.0,\newline 23.1 & 1.4, 1.3, 1.2,\newline 1.0, 1.0, 1.0,\newline 1.2\\
\enddata 
\tablenotetext{a}{Dark Energy Survey \citep{des16}; \url{https://www.darkenergysurvey.org}.}
\tablenotetext{b}{\cite{Dey2019}; \url{http://legacysurvey.org.}}
\tablenotetext{c}{DECam Legacy Survey; \url{http://legacysurvey.org/decamls/.}}
\tablenotetext{d}{Beijing-Arizona Sky Survey \citep{zou17}.}
\tablenotetext{e}{Mayall z-band Legacy Survey \citep{dey16}.}
\tablenotetext{f}{DECam Near Earth Object survey \citep{allen16}.}
\tablenotetext{g}{DECam Plane Survey \citep{schlafly18}; \url{http://decaps.skymaps.info.}}
\tablenotetext{h}{Blanco DECam Galactic Bulge Survey \citep{rich15}, \citep{johnson20}.}
\tablenotetext{i}{Deep Synoptic Study of the Galactic Bulge.}
\tablenotetext{j}{Light Echoes of Galactic Explosions \citep{rest15}.}
\tablenotetext{k}{ Survey of the Magellanic Stellar History \citep{Nidever2017}; \url{http://datalab.noao.edu/smash/smash.php.}}
\tablenotetext{l}{Blanco Imaging of the Southern Sky.}
\end{deluxetable*}

The NSC data are publicly available through the Astro Data Lab\footnote{\url{https://datalab.noirlab.edu}}, and can by accessed via SQL queries.  The catalog will be updated periodically, with the second data release containing over 68 billion measurements of 3.9 billion objects \citep{Nidever2020}.  Figure \ref{fig:nsc_data} shows the object density \textit{(left)} and exposure coverage \textit{(right)} of NSC DR1.  \editoneb{An overview of NSC DR1 can be found on the Astro Data Lab site\footnote{\url{https://datalab.noirlab.edu/nscdr1/index.php}}, and a complete description is available in \cite{Nid2018}.}

\editoneb{From the NSC DR1 exposures table, the average exposure time is 96.28 s, with a minimum of 0.01 s and a maximum of 1300 s.}  \edittwob{Figure \ref{fig:exp_hist} shows the number density for the full range of exposure times, broken down by instrument.  As described in Section \ref{ssec:hp_check}, the NSC is only searched for SSOs in areas covered by at least 3 exposures; the difference between the total exposure list and the searched exposure list is also shown in Fig. \ref{fig:exp_hist}.  All exposures are run through the NOIRLab Community Pipeline.  \cite{flaugher} report a shutter accuracy of $\leq$1ms for DECam, which provides both the largest sample of exposures and the widest range of exposure times.  Images with long exposure times may contain detections of faint TNOs detected as point sources as well as streaking from SSOs exhibiting higher proper motion; for an exposure time of 1300 seconds, an object moving $\mu$=20\arcsec relative to the background would leave a $\backsim$7.2\arcsec trail, within the $\backsim$1080\arcsec range of a typical single DECam CCD (2048x4096 pixels with pixel scale 0.26\ethreesout{37}).}

\editoneb{The NSC measurements table (\texttt{nsc\textunderscore dr1.meas}) provides the main souce of data for this work.  Each source-extracted measurement is represented as a row of data in the table, with a total of 34,658,213,888 rows.  31 columns of detection parameters include position and positional uncertainty (\texttt{RA, Dec}), exposure filter, apparent magnitude (\texttt{mag\textunderscore auto}), photometry information, and the observation Modified Julian Date (\texttt{MJD}).   An overview of all columns of the measurements table table is also available at Astro Data Lab\footnote{\url{https://datalab.noirlab.edu/query.php?name=nsc_dr1.meas}}.}

Use of the NSC may provide solutions to some of the limitations common to SSO searches described in the previous section.  One issue is the potential for observational bias in the regions covered.  The focus of surveys is often away from the galactic plane, due to the higher density of objects in that area\editoneb{ polluting the recovered SSO population}.  NSC DR1 covers 3/4 of the sky and approximately half of the galactic plane is included in the observed region, meaning this project searches portions of the sky that are under-examined for SSOs. 

\editoneb{A \etwosout{fall}drawback of the NSC is that exposures of the most dense areas are often oversaturated, resulting in a lack of distinction between real sources and image artifacts, and therefore a higher rate of false positives in SSO searches near the galactic plane ($\delta \leq 15^{\circ}$) and the Magellanic Clouds (5$^{\circ}$ around LMC and 3$^{\circ}$ around SMC).  The NOIRLab Community Pipeline used in the NSC calibration process masks saturated stars and their bleed trails, however some effects from bright saturated star still get through including scattered light in the periphery.  These generally do not line up in any preferred direction, and their effect must be accounted for in the validation of any tracklets found in the NSC. } 

The limiting magnitude of a survey is also very important: distant SSOs are quite faint and will be undetectable to most telescopes during the bulk of their orbits due to low albedo or large distance.  The depth of most NSC exposures is approximately 23rd mag, an adequate value for detecting TNOs whose magnitudes are typically fainter than 20th mag. The NSC also contains multi-band information that provide colors of objects and will shed further light on their properties.  In a study of the colors of almost 100 TNOs, \cite{schwamb19} found two taxonomic groups, and the NSC should contain measurements of enough TNOs for a similar investigation (see  Sec. \ref{sec:results}).

Any newly discovered objects may also clarify the size distribution of solar system bodies.  For instance, \cite{Shep2010} showed a lack of discovered intermediate and small Neptune Trojans.  Missing objects may be uncovered in the NSC after orbit determination, or may otherwise support these missing Trojans.  By investigating the range of asteroid sizes, new parameters can be established regarding the accretion and formation history of the solar system.

Finally, the NSC covers much of the southern celestial hemisphere, which has been largely under-searched in SSO surveys.  \editoneb{Five years of observations indicates the likelihood of orbiting bodies appearing in the NSC multiple times.   An issue is the intermittent observational pattern; certain areas of the NSC have insufficient exposure coverage for SSO detection, as the contributing surveys were designed for various purposes.  At least three exposures of a field are required to detect tracklets as defined in this work.  Section \ref{ssec:hp_check} describes the selection of areas with adequate NSC coverage.}

Because of these reasons, the deep, multi-band, time-series NSC information will allow this work to markedly improve the census of solar system bodies and help our understanding of planet formation.  Detections of Planet 9 may be hiding in the NSC, and, at the very least, any KBOs found in the catalog will provide further investigation into its proposed existence.



\begin{figure*}[ht]
\centering
\includegraphics[width=1\hsize]{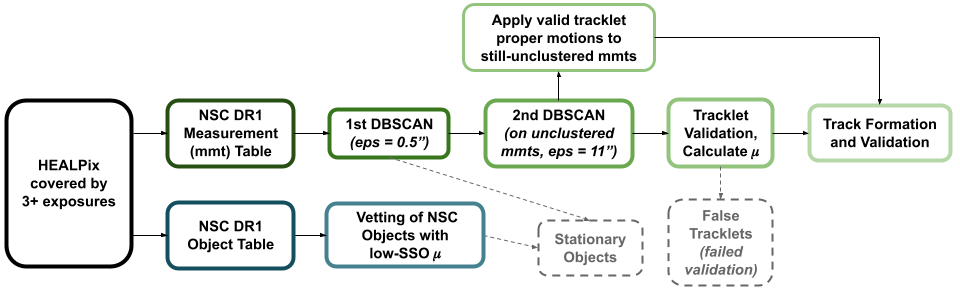}
\caption{CANFind flowchart, beginning with HEALPix covered by 3+ exposures \textit{(left)}.  The top row follows the main CANFind process from the NSC Measurement Table to Track Formation \& Vaidation.  The bottom row indicates the NSC objects that were analyzed as potential SSOs, referred to as ``Vetting NSC Objects''.}
\label{fig:flow}
\end{figure*}

\section{Moving Object Detection Method}
\label{sec:meth}

Here we describe the CANFind method used to detect SSO moving objects in the NSC catalog.  \editoneb{The source code is written in Python and uses the  \texttt{scikit-learn}\footnote{\url{https://scikit-learn.org/}} \citep{scikit-learn} machine learning package.  CANFind relies on two \texttt{scikit-learn} algorithms: the Density-Based Spatial Clustering of Applications with Noise \citep[DBSCAN;][]{ester96} and RANdom SAmple Consensus \citep[RANSAC;][]{fischler}, a linear regressor that is robust to outliers.}  Beginning with the NSC \texttt{meas} table of individual exposure measurements, the main steps are:
\begin{enumerate}
    \item Initial spatial clustering with DBSCAN (eps = 0.5\arcsec) to detect and remove stationary objects and initial detection of moving objects with moderate angular velocities.
    \item Second spatial clustering with DBSCAN (eps = 11\arcsec) to identify detections of SSOs which form tracklets.
    \item Validation of tracklets obtained in steps 1 \& 2 using robust linear regression (RANSAC) and the Pearson Correlation Coefficient (PCC).  Proper motions of valid tracklets are measured.
    \item Track formation and validation by linking tracklets and unclustered measurements by using proper motion to predict object position at common epochs.
\end{enumerate}

A flowchart outlining the process described in this section can be seen in Fig. \ref{fig:flow}.

\subsection{Spatial Clustering with DBSCAN on Hyalite}
\label{ssec:dbscan}
\eonesout{CANFind uses the Density-Based Spatial Clustering of Applications with Noise \citep[DBSCAN;][]{ester96}} \editoneb{DBSCAN is used} in an iterative process to assign measurements to unique objects, both stationary and moving.  \etwosout{We used the version of DBSCAN in the python \texttt{scikit-learn} \citep{scikit-learn} machine learning package.  }DBSCAN clusters data in areas of high density, and was developed in order to categorize the objects of a database into relevant groups. 

This particular algorithm was selected for a number of reasons, the most important being that minimal knowledge of the data set is required to establish the two DBSCAN input parameters set for this project; the eps-neighborhood (\texttt{eps}) of a point, which is the maximum distance between two points for one to be considered as in the neighborhood of the other, and \texttt{n}$_{points}$, the minimum number of points required to define a cluster.  In the case of this work, DBSCAN is applied to measurements using only the information of spatial location, and \texttt{eps} and \texttt{n}$_{points}$ are determined based on the type of object being clustered.  DBSCAN also does not require a set number of priors, which is advantageous as there is no way of knowing how many objects, stationary or otherwise, will be in any area of the sky.  \editoneb{One caveat is that DBSCAN does not take into account the positional uncertainties of measurements; they are, however, included in the calculation of tracklet proper motion.  }

The second reason for the selection of DBSCAN is its ability to identify clusters with arbitrary shape.  This allows one clustering method to be applied to all types of objects in the NSC identified in this work.  It can cluster the measurements of both stationary objects (SOs), which lie in close proximity to each other in a roughly spherical pattern, as well as the more linearly-distributed measurements of moving objects.  \editoneb{As DBSCAN is only used to connect intra-night detections, more complex SSO motion can be ignored for the extent of its use.}  Furthermore, the capability of DBSCAN to identify outlier points is essential to the method of application, as described in the following sections.  \editoneb{Some of \texttt{scikit-learn}'s other clustering algorithms offer this feature; for example t}he Variational Bayesian Gaussian Mixture Model (VBGMM\eonesout{, also available at \texttt{scikit-learn}}) was also considered as it can identify clusters with both strong and weak linear correlations.  Ultimately, the number of necessary parameters and the slow computation speeds (this algorithm is not good for very large datasets) eliminated VBGMM in favor of DBSCAN.  

Lastly, DBSCAN is very efficient for large databases with significantly more than several thousand objects \citep{ester96}.  As we analyze billions of measurements, it is essential that CANFind is as efficient as possible to maintain reasonable computation times.  This is aided by Montana State University's Hyalite computer cluster, enabling the application of these methods on the vast extent of the NSC through a parallelized framework.  Hyalite has 1,300 cores and 77 computing nodes with either 64 or 256 GB of RAM.  By running CANFind on the Hyalite cluster, many areas of the sky can be searched simultaneously and the computation time is dramatically reduced.

\eonesout{Analysis is run on one}\editoneb{We take advantage of the NSC coming pre-divided into} HEALPix\footnote{\url{https://HEALPix.jpl.nasa.gov/}} \citep[\editoneb{sections of equal area,}][]{Gorski2005} \eonesout{at a time }with \editoneb{NSIDE=265 indexed in the \texttt{ring265} column of the object table, and query 4 neighboring HEALPix at a time to combine them into an area that takes CANFind an average of 5 seconds to analyze, depending on the number of sources and measurements present in the field.  The resulting combined HEALPix, with }NSIDE=128 for a resolution of approximately 0.5$^{\circ}$\editoneb{, is run through the CANFind process with several of its neighboring HEALPix as a job on Hyalite's \texttt{slurm} queue}.  Although Hyalite allows several HEALPix to be processed simultaneously, computation speed is hampered slightly by \editoneb{PostgreSQL} query limitations\eonesout{ from Astro Data Lab}.  \editoneb{Fewer HEALPix can be analyzed per job in fields with high object density, as this increases the query processing time and slows down the job's progression. Additionally, jobs had to be submitted in incremental batches as the database can only handle a limited amount of queries at a time.  Depending on the field density, 1--10 jobs were submitted every 10--60 minutes for 1--5 HEALPix per job. Using roughly 10 nodes at a time with 32 or 40 CPU/node, the NSC took approximately one week to analyze on Hyalite.}

\begin{figure*}[t]
\centering
\includegraphics[width=1.0\hsize,angle=0]{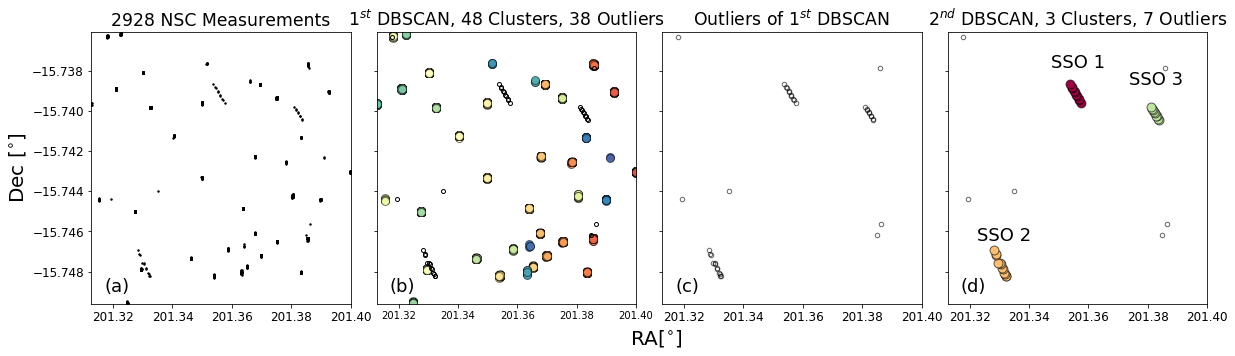}
\caption{An example spatial map of NSC measurements showing clear detections of three Solar System Objects (SSO).  The SSOs move rapidly ($\sim$40\arh  \editoneb{or 0.27\ded}) and show a ``tracklet'' while background objects keep a fixed position ({\em panel a}).  Measurements belonging to fixed objects are clustered ({\em panel b}) and removed, leaving behind outlier measurements ({\em panel c}) to be assigned to moving objects.  A second iteration of clustering identifies the SSO tracklets ({\em panel d}).}
\label{fig:sso_demo}
\end{figure*}

The rest of this section describes how DBSCAN is used to assign NSC measurements to unique objects, visualized Figure \ref{fig:sso_demo}.  The first iteration of DBSCAN identifies and removes stationary objects moving less than 2.5\arh \editoneb{(0.017\ded)} for a separate vetting process, leaving remaining measurements to be clustered into distinct moving objects during the second iteration.  These data will be fit to solar system orbits to create object candidates, and the properties of these candidates will be derived and analyzed.

\subsection{Selecting areas of the sky for initial tracklet search}
\label{ssec:hp_check}
The NSC has a very high object density within the Galactic plane of the Milky Way. Exposures from the galactic plane also contain a high number of very bright and saturated stars which can cause various artifacts in the source catalog.  Because these artifacts heavily impair the performance of the DBSCAN detection method, fewer objects are expected to be extracted within $\pm15^{\circ}$ of the galactic plane.  \editoneb{This is exacerbated by the high density of background objects in both the galactic plane and the Magellanic Clouds, as crowding hampers the ability of DBSCAN to differentiate unique sources.  A sample of tracklets extracted from these areas by CANFind is examined in finer detail in Section \ref{sec:results}.}  \eonesout{Due to the high object density characteristic of the Magellanic Clouds, data within $5^{\circ}$ of the Large Magellanic Cloud (LMC) center or within $3^{\circ}$ of the Small Magellanic Cloud (SMC) center were omitted from the search.} 

\begin{figure}
\centering
\includegraphics[width=1.0\hsize,angle=0]{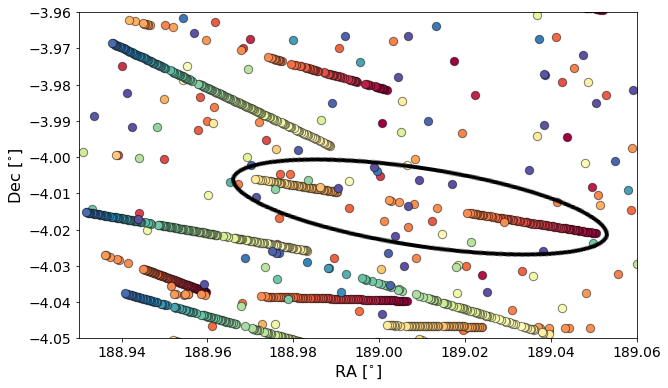}
\caption{An example of track formation, with color as time of \editoneb{observation (over two nights, MJD=56746-56747)}.  The two circled tracklets, along with various unclustered measurements between them, are linked together to form a track using the position prediction method described in subsection \ref{ssec:mo}.}
\label{track_form}
\end{figure}

We used the HEALPix scheme with nside=128 to break up the sky into smaller chunks to be analyzed separately. For the initial search, only areas of sky with adequate cadence to capture sequential displacements of the faster-moving SSOs were used.  The selection criteria was that the HEALPix must have coverage from at least three or more exposures taken within roughly 20 minutes of each other.  This reduced the 153,472 HEALPix covered by the NSC to 132,884 to be studied with CANFind.

\subsection{Vetting NSC Objects}
\label{ssec:nscobj}
Because some SSOs have proper motions low enough to be represented in the NSC object table, we run an additional test on all NSC objects with at least 3 detections and a recorded NSC proper motion above \editoneb{0.0041\ethreesout{34}\arh (}2.7\ethreesout{56}e-5\ded\editoneb{)}.  This value corresponds to the average proper motion of 2015 TG387, a TNO with the second-largest semimajor axis known to date \editoneb{(1085 AU)} and an orbit stable over the history of the solar system \citep{shep19}.  

\eonesout{Spatial error becomes more significant in the validation of these potential SSOs due to the measurements of.  As a result, this validation process is more lenient in terms of linearity so PCC is not used.}  \editoneb{Every object above the minimum proper motion is analyzed to estimate the likelihood that it is an SSO.}  \eonesout{RANSAC is still} \editoneb{With a set residual threshold of 0.001$^{\circ}$, RANSAC} is applied to the spacial and temporal object data to remove galaxies with several NSC detections over multiple years.  The \texttt{curve$\_$fit} function from \texttt{scipy.optimize} is applied to the measurements of each object to calculate proper motion in RA and Dec [\arh].  

A signal to noise cut of S/N$\geq$4 was made on net proper motion in order to reduce the contamination of stars and galaxies among true detections of SSOs.  To further reduce this contamination, a cut on the reduced chi squared value was made, keeping all objects with $\chi_{\nu}^2\leq1.5$.  \editoneb{Despite quality cuts, it is expected that a greater number of false tracklets are present in t}he 2095 objects that passed this validation step\editoneb{. Nonetheless, they} are considered to be tracks or tracklets, and\eonesout{,} therefore\eonesout{,} will be tested as SSO candidates with the rest of the tracklets detected by CANFind.

\begin{figure*}[ht]
\centering
\includegraphics[width=1\hsize,angle=0]{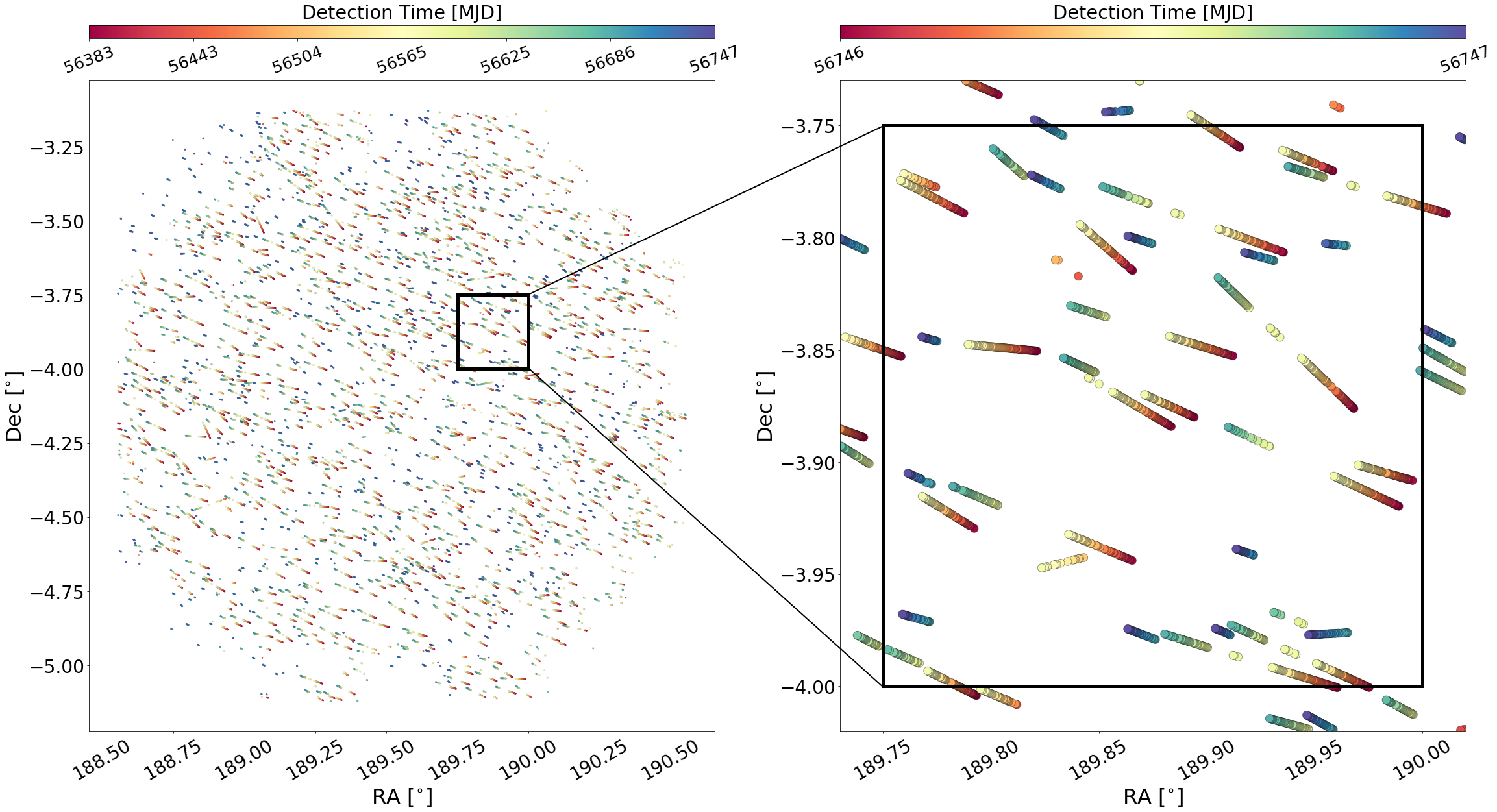}
\caption{\editoneb{2,941 t}racklets, color-coded by \editoneb{measurement} detection time, in one DECam field in the ecliptic plane.  \editoneb{Field depth goes to 25.7 mag.}}
\label{fig:decam_field}
\end{figure*}

\subsection{Stationary Object Identification}
\label{ssec:so}

DBSCAN is first used to identify measurements belonging to slow-moving or stationary objects using a relatively small \texttt{eps} = 0.5\arcsec and \texttt{n}$_{points}$ = 2.  The \texttt{eps} value was selected to reflect the original object identification of the NSC.  The NSC \texttt{object} table was created by associating points within 0.5\arcsec of each other, but this did not account for the faster proper motion ($\mu$) of most SSOs.  For this reason, only the NSC measurement table was used for this step.  Panels {\em a} and {\em b} of Figure \ref{fig:sso_demo} demonstrate the application of DBSCAN on 2,928 measurements from the NSC in HEALPix pixel 124958 positioned a few degrees from the ecliptic plane, forming 48 clusters (SOs) and leaving 38 outliers.  The proper motions of \editoneb{many of }these objects are already given by the NSC catalog and \eonesout{will be used}\editoneb{were referenced in the analysis} for objects exhibiting Solar System-like motions.  The DBSCAN outliers are retained to be assigned to faster-moving objects in the next step.  Because the \editoneb{source-extracted} measurements of \eonesout{each clustered object must be very closely spaced, nearby} resolved galaxies \editoneb{may be more than 0.5\arcsec apart, nearby galaxies} are often not identified in this first step, and \eonesout{are dealt with}\editoneb{must be removed }later during the validation process.

\subsection{Identifying Tracklets and Tracks}
\label{ssec:mo}
The identification of tracklets and tracks is performed in three steps:

{\bf Linking Measurements into Tracklets:} Once measurements of SOs are identified and removed, the remaining measurements are assigned to unique moving objects by the second iteration of DBSCAN.  When closely-spaced exposures ($\sim$20 minutes apart in this example) are analyzed, measurements of fast moving objects form spatially linear structures called ``tracklets''.  Because these measurements are farther apart from each other than those belonging to SOs, a new \texttt{eps} of 11\arcsec is used, as seen in panel {\em d} of Figure \ref{fig:sso_demo}.  Each DBSCAN cluster must now contain \texttt{n}$_{points} \geq 3$ detections, according to our definition of a tracklet as a collection of at least three displaced detections of a unique object.  Therefore, each cluster is a tracklet that may represent the detections of an SSO, and are referred to as \textit{``CANFind tracklets''}.

{\bf Tracklet Validation:} Each clustered tracklet undergoes a validation process to ensure that its measurements are consistently linear in space and time, and to eliminate any falsely clustered data.  \eonesout{As previously mentioned, s}\editoneb{S}ome \editoneb{nearby, }spatially-resolved galaxies \eonesout{are cataloged as separate objects in the NSC} with a roughly linear \editoneb{distribution of measurements}\eonesout{shape of detections, so they} are mistakenly clustered as tracklets\editoneb{ in this step, as DBSCAN often fails to identify them as stationary objects if that distribution is too large}.  However, they are easily identified through examination of their temporal information.  In the creation of the NSC, multiple source detections from a single galaxy were sometimes extracted from one exposure, so resulting DBSCAN clusters do not produce a realistic trajectory for an SSO.

To test each tracklet, its measurements are analyzed using \editoneb{RANSAC} \eonesout{the Random Sample Consensus \citep[RANSAC;][]{fischler}, a linear regressor that is robust to outliers} \editoneb{on both the spatial coordinates and the tracklet's normalized (\textit{position,time}) information}.   Any outlier found by RANSAC is likely an incorrectly clustered measurement, and is eliminated from the tracklet.  As an additional step, the Pearson Correlation Coefficient (PCC) is calculated for each tracklet.  \edittwob{The assumption that} a true tracklet has measurements that fall into a straight line \edittwob{motivates the PCC cutoff value of $\geq|0.9|$ used by CANFind.  }\etwosout{, and the PCC gives a measure of how tightly correlated its points are}.  If the spread \edittwob{of tracklet measurements} is too large, then the cluster likely only contains measurements of a galaxy, not an SSO.  This validation process disposes of many erroneous tracklets, making the next step of linking tracklets together a less computationally expensive process.

{\bf Linking Measurements and Tracklets into Tracks:} \editoneb{Track formation in CANFind is meant to associate tracklets with both other tracklets and unclustered measurements left from the previous clustering steps.  }After a tracklet is verified,
the slope values determined by RANSAC are saved as the proper motion in RA and Dec [\ded].  Using these values, a position-prediction method is applied to the tracklets and remaining unclustered measurements within the HEALPix, allowing them to be connected into longer ``tracks'', which consist of at least two tracklets \editoneb{and typically contain detections from multiple nights.  As the assumption of linear motion starts to fail for trajectories spanning so much time, the track validation step is less \etwosout{rigorous}\edittwob{restrictive} than tracklet} \edittwob{validation; that is to say, the positional uncertainty allowed for predicted positions of tracklet measurements is large (0.005$^{\circ}$) compared to measurement uncertainty to account for the deviation from linearity that SSOs display over multiple nights.}

\begin{figure*}[t]
\centering
\includegraphics[width=1.0\hsize,angle=0]{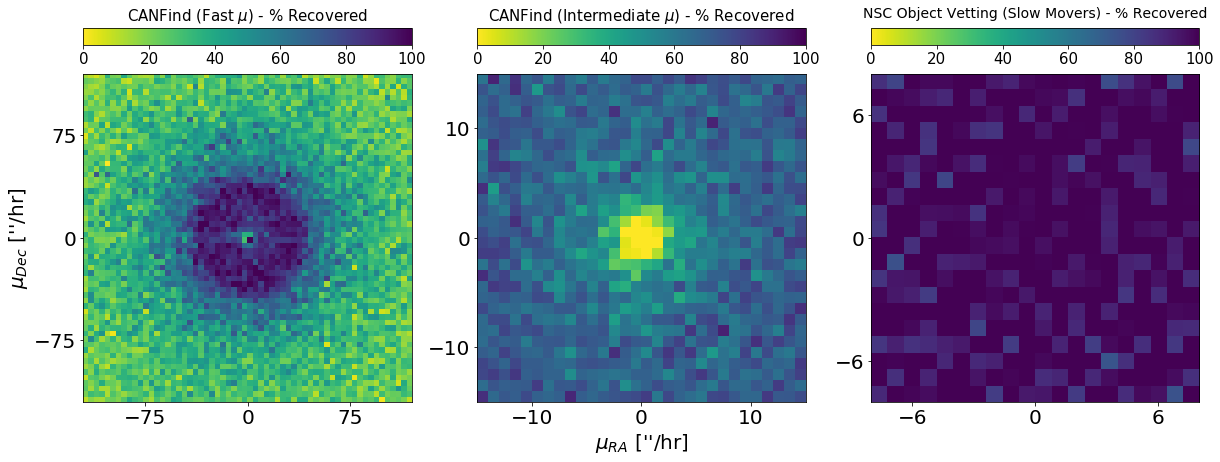}
\caption{Density plots from the simulation showing 3 general ranges of tracklet proper motion $\mu$ [\arh]. CANFind results show the algorithm's upper \textit{(left)} and lower \textit{(center)} limits set by the parameters used for DBSCAN.  CANFind leaves a gap around $\mu$=0 of about 2.5\arh \editoneb{(0.017\ded)} in radius, which is filled in by the NSC object vetting process \textit{(right)} described in Sec \ref{ssec:nscobj} .}
\label{fig:simdiffs}
\end{figure*}

The proper motion of a tracklet is used to project its position to the observation time of other tracklets and unclustered measurements.  This is done twice - first, the calculated proper motion of a single tracklet is applied to all other tracklets and unclustered measurements in the healpix, and all points are moved to the location they would hypothetically be at a common time.  If the positions coincide within a few times their measurement uncertainties, then the tracklets and unassigned measurements are linked together into tracks.  This is repeated for every tracklet in the healpix.  A second ``projection'' process is applied for comparison, this time with each tracklet moving at its own calculated velocity to a common epoch.  An example of track formation is shown in Figure \ref{track_form}.

To demonstrate the typical distribution of tracklets within the ecliptic plane, Figure \ref{fig:decam_field} roughly covers the area of one DECam field, with color as detection time.



\section{Efficiency Simulations} 
\label{sec:sim}

To test the efficiency and completeness of CANFind\editoneb{ and the NSC Object Vetting process to detect moving objects,} a simulation was created that generated synthetic tracklets with \eonesout{various}\editoneb{linear trajectories and a range of} proper motions \editoneb{that are} characteristic of SSOs.  \editoneb{These tracklets had between 3 and 10 measurements, and the average tracklet cadence was between 20 minutes (CANFind) up to several hours (NSC Object Vetting).  Measurements were given equatorial coordinates, an observation date, and a magnitude between 14--25 with an uncertainty (\texttt{magerr\textunderscore auto}) that was then used to calculate the positional error in (\textit{RA,Dec}).  }These generated tracklets were separately added to un-analyzed NSC measurements from a HEALPix (nside=128) in a field of average object density.  \eonesout{Both CANFind and the NSC Object Vetting process were subjected to simulations.}  

\editoneb{The first set of tests varied the range of input tracklet proper motion with the goal of determining biases set by CANFind's use of DBSCAN.  }Simulations of 100,000 tracklets each were run for different ranges of $\mu$ to explore the upper and lower limits of CANFind's range.  Results for $0\leq \mu\leq$\eonesout{200}\editoneb{ 100\arh (0.67\ded), $0\leq\mu\leq$ 15\arh (0.1\ded),} and $0\leq\mu\leq$\eonesout{20} \editoneb{8\arh} \editoneb{(0.053\ded)} are shown in Figure \ref{fig:simdiffs}.  

\editoneb{The performance of CANFind for a range of magnitude uncertainties was explored in a second set of tests using an average FWHM value = 1.1\ethree{8}\ethreesout{7734} from the NSC exposure table.  The range of \texttt{magerr\textunderscore auto} was sampled from several areas including the galactic plane, ecliptic plane, and in a field away from both planes.  Between simulations of 100,000 tracklets each for three ranges of magnitude uncertainties (0.0002--0.002 mag, 0.002--0.02 mag, and 0.02--0.2 mag) CANFind's detection efficiency only changed slightly (97.6\%, 97.2\%, and 95.5\% for the three ranges, respectively).}

\editoneb{In addition to the 600,000 tracklets tested over varying $\mu$ and \texttt{magerr\textunderscore auto}, a cumulative 200,000 tracklet test over full proper motion and uncertainty ranges found that} \eonesout{The detection efficiency test of}CANFind recovered 96.5\% of synthetic tracklets \editoneb{with average measurement spacing ($\Delta s$) between 0.5--10.9\arcsec for}\eonesout{with} proper motions less than 1.\ethree{4}\ethreesout{397}\ded (209\ethreesout{.6}\arh) to within 1\arh of their input proper motion. The limits of CANFind are a result of the parameters used in both instances of DBSCAN.  The first iteration (\texttt{eps}=0.5\arcsec) to dispose of stationary objects sets the lower limit, (center panel of Fig \ref{fig:simdiffs}), and the upper limit (left panel of Fig. \ref{fig:simdiffs}) is set by the second iteration (\texttt{eps}=11\arcsec).  As a result, this search optimizes CANFind to detect tracklets with cadences between \eonesout{2}\editoneb{0}.5--11\arcsec \editoneb{(0.00012--0.003$^{\circ}$)}.   

\editoneb{CANFind's false positive rate was also tested in fields of varying object density.  Fake, linear sky-plane tracklets were created with measurements spanning one night or less, and given random, realistic positional uncertainties and observation times.  One set of 1000,000 synthetic ``fake'' tracklets was analyzed in a HEALPix 5$^{\circ}$ above the galactic plane (35,993 objects, 159,191 exposures) and in a HEALPix 6$^{\circ}$ above the ecliptic (7,990 objects, 248,004 measurements).  The fake tracklets, like the ``true'' tracklets created for previous simulations, consisted of 3 to 10 measurements spaced less than 0.003$^{\circ}$ (10.8\arcsec) from each other.  In both fields, 3\% of fake tracklets were falsely recovered by CANFind.}

\editoneb{Finally, the efficiency of the track linking process was tested at a range of velocities between $0\leq\mu\leq$60\arh (0.\etwosout{1}\edittwob{4}\ded).  100,000 tracks were created for this test by extending a synthetic tracklet beyond the timeframe of one night, either creating a new tracklet or adding unlinked "detections" of the object "observed" at a later time.  Track measurements were given the same information given in tracklet simulations, and all complete tracklets created to form tracks were correctly recovered by CANFind.  80\% of the synthetic tracks were completely identified by CANFind.}

\section{Results}
\label{sec:results}

\begin{figure}[t]
\centering
\includegraphics[width=1\hsize]{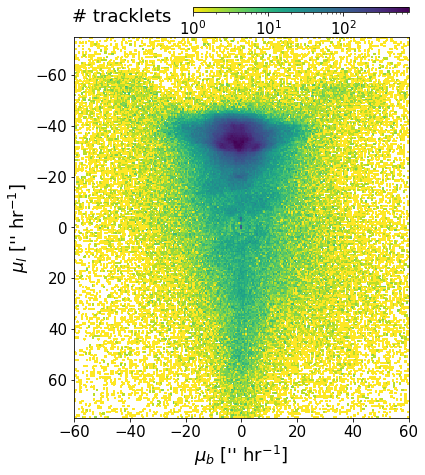}
\caption{Density plot of the proper motions [\arh] in ecliptic coordinates of tracklets found in the NSC using CANFind.  Distinct groups of families are present, most distinctly the main belt group ($\mu_{l}\sim$-40\arh \editoneb{or -0.27\ded}), Jupiter Trojans ($\sim$-20\arh \editoneb{or -0.13\ded}), and some Kuiper Belt Objects ($\sim$-3\arh \editoneb{or 0.02\ded}).}
\label{fig:proper_motion}
\end{figure}

\begin{figure}[t]
\centering
\includegraphics[width=1\hsize,angle=0]{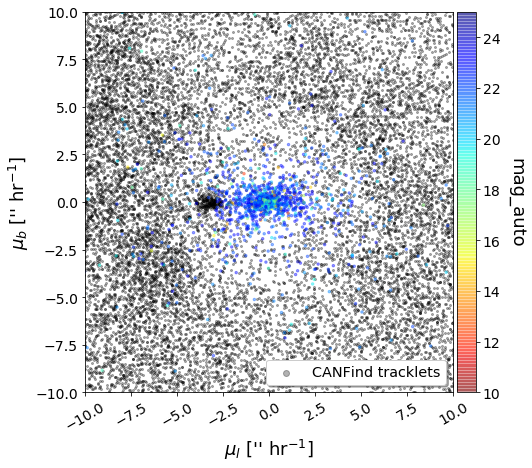}
\caption{Proper motions of tracklets detected with the main CANFind algorithm (gray), and objects identified by the vetting process (sec. \ref{ssec:so}) with color as apparent magnitude.  The gap around $\mu$=0 left by CANFind (see center panel of Fig. \ref{fig:simdiffs}) is filled in by this extra step.}
\label{fig:nsc_obj}
\end{figure}

\begin{figure*}
    \centering
    \includegraphics[trim={1cm 0cm 0cm 0cm},width=0.48\hsize,angle=0]{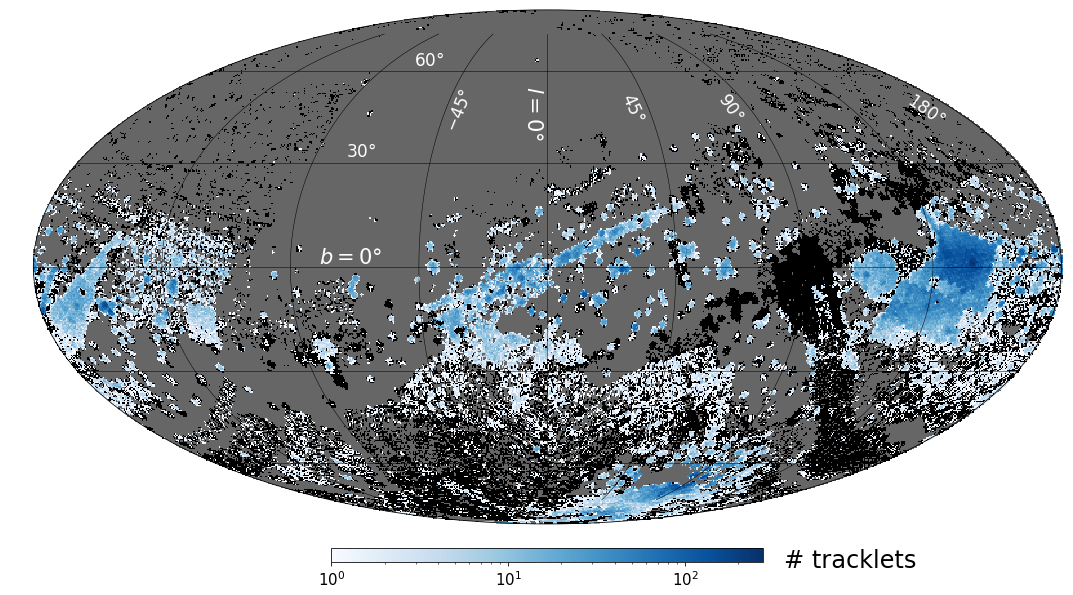}
    \includegraphics[trim={1cm 0cm 0cm 0cm},clip,width=0.48\hsize,angle=0]{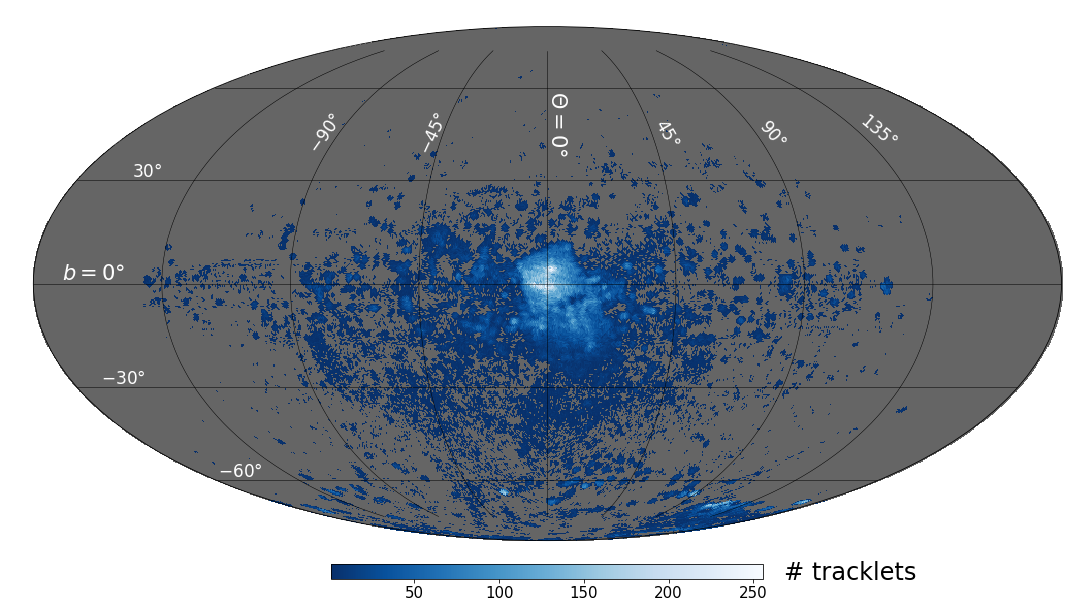}
    \includegraphics[trim={1cm 0cm 0cm 0cm},clip,width=0.48\hsize,angle=0]{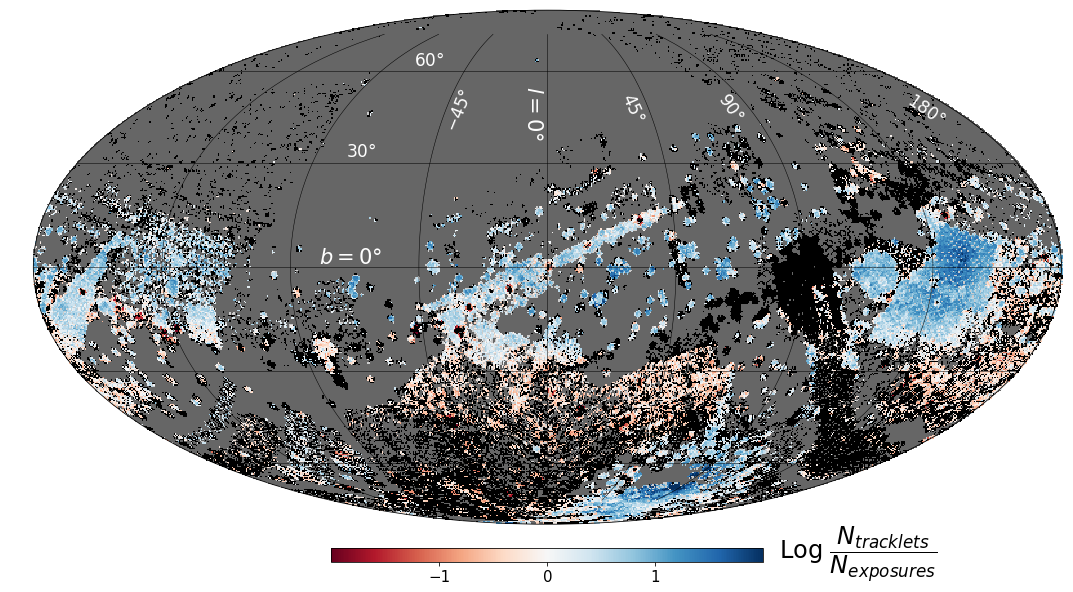}
    \includegraphics[trim={1cm 0cm 0cm 0cm},clip,width=0.48\hsize,angle=0]{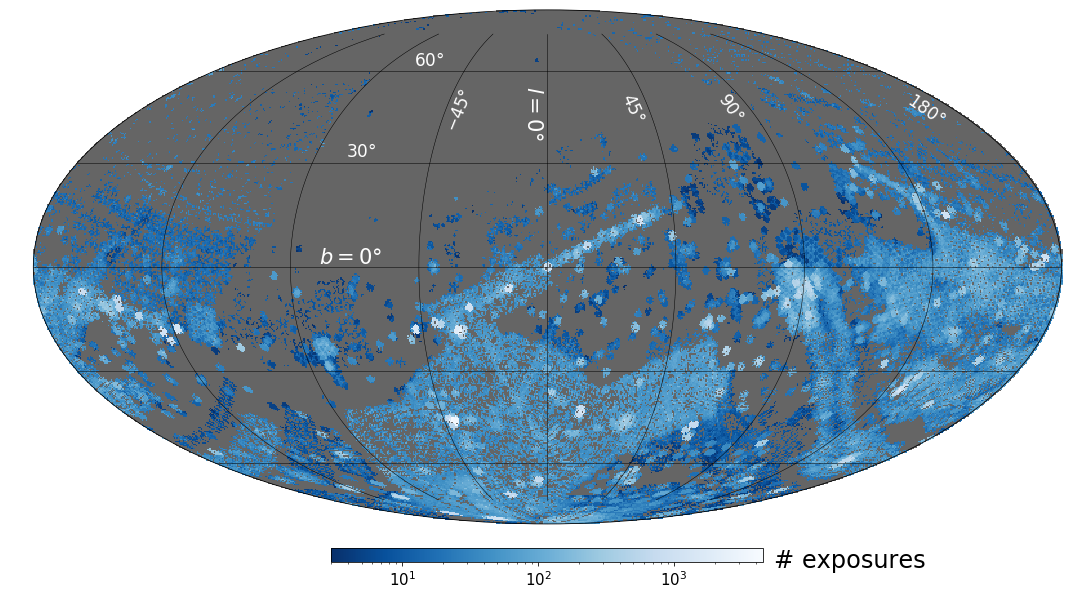}
    \caption{\textit{(upper \editoneb{left})} \eonesout{Density plot}\editoneb{Tracklet density} (number of tracklets per HEALPix, NSIDE=128) of NSC DR1 in ecliptic coordinates; \editoneb{grey areas had insufficient coverage to be searched, black areas were searched but contained no tracklets.  \textit{(upper right)} Tracklet density in opposition-centric ecliptic coordinates.  In place of ecliptic longitude, the x-axis represents $\Theta$ = angular separation from opposition, which is 180$^{\circ}$ from the apparent ecliptic longitude of the sun at the time of tracklet observation.}  \ethree{Tracklets found at high ecliptic latitudes (i.e. LMC, SMC) are more likely to be false positives (see Sec. \ref{sec:results}).}  \textit{(lower \editoneb{right})} Tracklet\eonesout{ distribution}\editoneb{Exposure density }in ecliptic coordinates\eonesout{ with color as proper motion [\arh]}\editoneb{\textit{(lower left)} Density of the normalized tracklet/exposure ratio in ecliptic coordinates compares the relative amount of tracklet detection in an area to the amount of exposure coverage.}}
    \label{fig:nsc_tracklets}
\end{figure*}

\begin{figure*} 
    \centering
    \includegraphics[clip,scale=0.52]{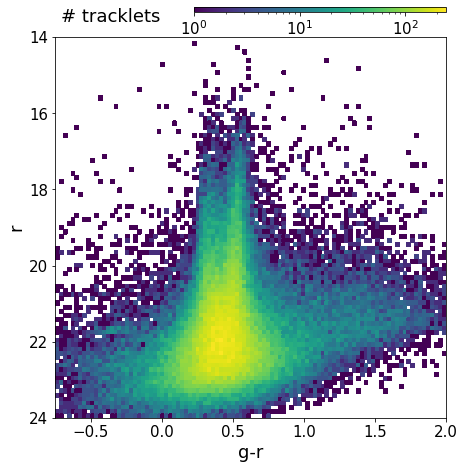}
    \hspace{0.5cm}
    \includegraphics[clip,scale=0.35]{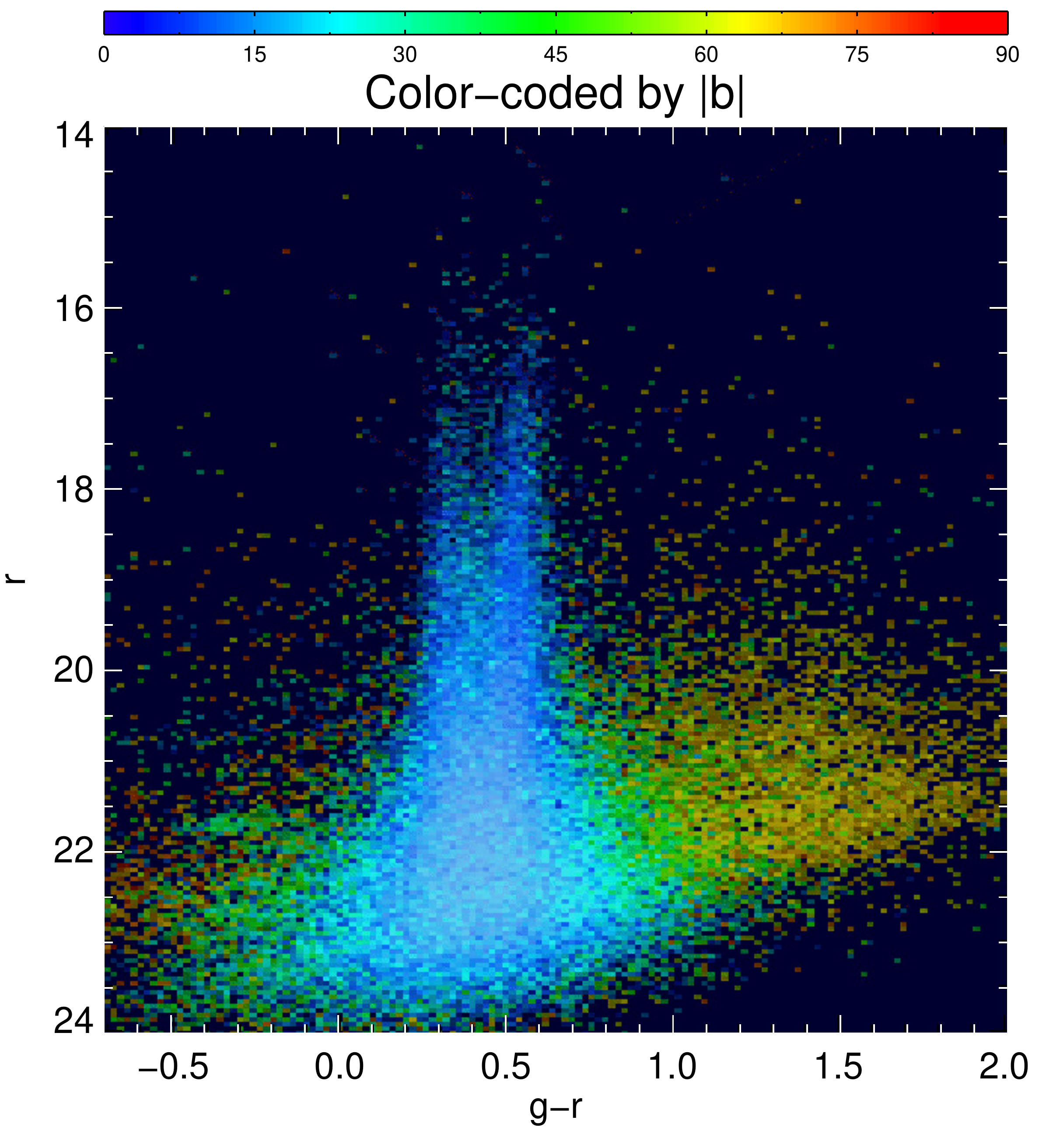}
    \caption{Color magnitude diagrams of the 107,604 tracklets found by CANFind with \textit{g} and \textit{r} band photometric data.  \textit{(left)} shows color as density, and \textit{(right)} shows saturation as density and color as angular distance from the ecliptic (\textpipe\textit{b}\textpipe, [$^{\circ}$]).  Tracklets that are far from the ecliptic tend to be high in proper motion and do not follow the bimodal color profile characteristic of most SSOs.}
    \label{fig:cmd}
\end{figure*}

The CANFind search for SSOs yielded \eonesout{524,501}\editoneb{527,055} tracklets in \eonesout{35,644}\editoneb{35,710} HEALPix of nside=128 covering \eonesout{27,882}\editoneb{29,971} square degrees of the sky.  The density of tracklets in ecliptic proper motion can be seen in Figure \ref{fig:proper_motion}.  The 2,094 tracklets detected by the NSC object vetting process fill the range between 0$\leq\mu\leq$ 2.5\arh \editoneb{(0.017\ded)} left by CANFind (Figure \ref{fig:nsc_obj}).  The distribution of tracklets in ecliptic coordinates \eonesout{color-coded by their proper motion (\arh) }is shown in Figure \ref{fig:nsc_tracklets}.  \edittwob{177,949 tracklets (40\%) contain only three detections.  }170,584 tracklets were detected within 5\% of the ecliptic, and 380,189 have negative ecliptic longitudinal proper motions.  Tracklets detected at high ecliptic latitudes tend to be high in proper motion ($\backsim$100\arh\editoneb{, or 0.67\ded,} and above).  Figure \ref{fig:cmd} shows these objects following a color profile similar to stars and galaxies, suggesting a higher rate of false tracklets in regions further from the ecliptic. 

\editoneb{Samples of tracklets from high-density areas (galactic latitude $\leq$ 15$^{\circ}$, 5$^{\circ}$ around LMC, 3$^{\circ}$ around SMC) were examined in more detail.  Despite greater NSC exposure coverage in the galactic plane, CANFind detected more tracklets near the Magellanic Clouds.  A visual inspection of tracklets pulled from this area revealed a higher number of false positives detected near the MCs.  It is likely that fewer tracklets were found in the galactic plane due to the crowding of sources and the optimal spacing between exposures in the MCs. }

\eonesout{In all, 71.6\% of tracklets have a net $\mu$ value between 25 and 45\arh.}Several clear structures \editoneb{in proper motion space }can be seen in Fig. \ref{fig:proper_motion}, reflecting different \eonesout{families}\editoneb{groups} of SSOs.  Most tracklets have proper motions that indicate main-belt orbits with several other general asteroid groups represented as well.  Most prominent, centered at roughly $\mu_{l}$=$-$40\arh \editoneb{(-0.27\ded)}, $\mu_{b}$=0\arh in $\mu$-space is a group of tracklets consistent with MBA proper motion.  The presence of Jupiter Trojans and Trans-Neptunian objects are evident at $\mu_{l}$=$-$20\arh \editoneb{(-0.13\ded)} and $\mu_{b}$=3\arh \editoneb{(0.02\ded)}.  A noticeable feature of proper motion distribution are the ``wings'', which extend from the MBA group.  \cite{Terai13} describe these structures as a result of the variety of inclination and semimajor axes values of MBAs.  A more detailed analysis of each group will require orbital fitting (see Sec. \ref{sec:future}).

Tracklets were detected in all photometric bands included in the NSC ($ugrizY$ and {\em VR}). \editoneb{If a single tracklet contained multiple detections in the same band, their mean value was recorded as the tracklet's magnitude in that band.} Figure \ref{fig:mags} shows the normalized distribution of magnitudes for the \textit{ugr} bands extending to $\sim 24$ mag.  Table \ref{tab:bands} gives the number of tracklets with data in each band.  This overall distribution of tracklet detections in the various bands largely reflects the underlying distribution of exposures in the NSC, but may also be affected by the asteroid compositions.

\eonesout{The color-color diagram in Figure \ref{fig:ccd} shows the bimodality in SSO composition identified by \cite{ivezic01}.}  \editoneb{Table \ref{tab:multibands} lists the six most populous colors represented in the tracklet information.  Interestingly, NSC tracklets show a bimodality in color identified in \citet{ivezic01}, most notably in \textit{g-r} vs \textit{r-i} (see Figure \ref{fig:ccd}).}  Objects which fall in the redder group are mainly S-type rocky (siliceous) asteroids, while those in the bluer group are C-type (carbonaceous) asteroids. 45.5\% of tracklets fall roughly into the C-type group, with the remaining fraction closer to S-type.

\edittwob{A fraction of the tracklet measurements were found in images with exposure times long enough to produce a trailing effect, as no morphology cuts are made on tracklet measurements to remove elongated detections.  NSC measurements are found with Source Extractor \citep{bertin96}, which can identify objects of various shapes including stars, elongated galaxies, and streaked SSOs.  As a result, some trailed measurements of SSOs are identified and present in the NSC measurements table.  Figure \ref{fig:long_exptime} gives an example of trails left by a tracklet detected by CANFind.  For reference, 119 of the 255,454 NSC DR1 images have exposure times $\geq$ 1000 seconds, and 881 have exposure times $\geq$ 600 sec.  We find an approximate upper limit on the proper motion of CANFind tracklets containing measurements from these images, proportional to exposure time.  This corresponds a the limit on trailed detections represented as single source detections in the NSC.  Trails longer than $\backsim$13'' are not identified by CANFind as the source extraction process will incorrectly identify multiple sources from a single streak, all with the same observation date.  Future analysis will properly identify longer trails.}

\begin{deluxetable}{ll}
\caption{Number of NSC tracklets with \eonesout{band X information}\editoneb{at least one detection in a photometric band.}}
\label{tab:bands}
\tablecolumns{2}
\tablewidth{0pt}
\tablehead{
\colhead{band} & \colhead{\# tracklets}
}
\startdata
\textit{u} & 5,400\\
\textit{g} & 138,903\\
\textit{r} & 161,229\\
\textit{i} & 129,610\\
\textit{z} & 47,948\\
\textit{Y} & 7,570\\
\textit{VR} & 192,178\\
\enddata
\end{deluxetable}

\editone{
\begin{deluxetable}{ll|ll}
\caption{\editoneb{Six most populous colors, ``multicolor'' groups}}
\label{tab:multibands}
\tablecolumns{2}
\tablewidth{0pt}
\tablehead{
\colhead{color} & \colhead{\# tracklets} & \colhead{multicolor} & \colhead{\# tracklets}
}
\startdata
\textit{g-r} & 107,604 & \textit{g+r+i} & 50,222\\
\textit{r-i} & 70,557  & \textit{r+i+z} & 16,323\\
\textit{g-i} & 70,161  & \textit{g+r+z} & 9,518\\
\textit{i-z} & 30,706  & \textit{g+i+z} & 8,505\\
\textit{g-z} & 13,335  & \textit{i+z+Y} & 2,806\\
\textit{r-z} & 22,563  & \textit{r+i+Y} & 940\\ 
\enddata
\end{deluxetable}
}




\section{Future Work} 
\label{sec:future}

Recovered tracklet \editoneb{measurements}, along with measurements not associated with stationary objects, are stored in a PostgreSQL database at MSU for ease of access, similar to the one used to house the NSC in the Astro Data Lab.  In the future, we plan to run CANFind on the second data release of the NSC and the same techniques described in this paper will be applied to the new data.

According to \cite{weryk20}, the Isolated Tracklet File (ITF) of the Minor Planet Center, the organization responsible for the designation of all minor SSOs, holds over 16.6 billion unlinked measurements. Once further developed, CANFind may be applied to the ITF in hopes of reducing its volume.

The following steps will be taken next:



\begin{itemize}
\item \textit{Linking Dispersed Measurements of Moving Objects with the Hough Transform:} 
While DBSCAN effectively extracts tracklets with closely-spaced measurements, it does not identify those with a more disperse layout.  These detections will be connected via the Hough Transform approach, a feature extraction technique traditionally used to isolate features in an image \citep{Duda1972}, often straight lines or simple shapes.  The process is related to the Radon Transform \citep{Radon17} and works roughly as follows: Every NSC measurement will cast votes for all trajectory parameters $(\rho,\theta)$ consistent with that measurement and the votes from other measurements accumulated in a 2-D parameter/accumulator array.  Multiple measurements voting for the same line will suggest a candidate trajectory, identified by peaks in the accumulator array. 



The development of a modified version of the Hough Transform designed to work on data points rather than images is the subject of ongoing work.

\begin{figure}
\centering
\includegraphics[width=1\hsize,angle=0]{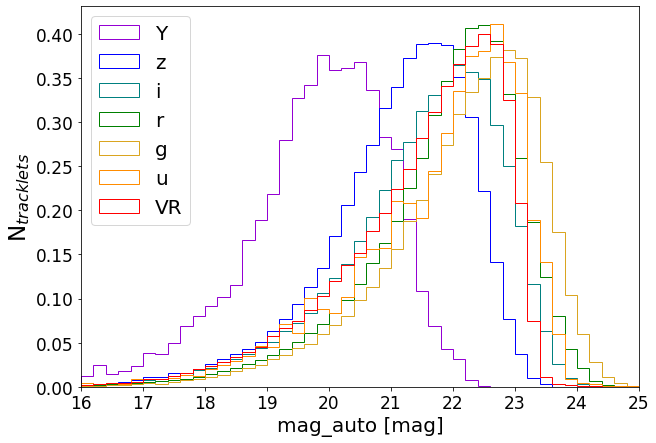}
\caption{Normalized histogram of the apparent magnitudes of NSC tracklets extracted by CANFind.}
\label{fig:mags}
\end{figure}

\begin{figure}
\centering
\includegraphics[width=1\hsize]{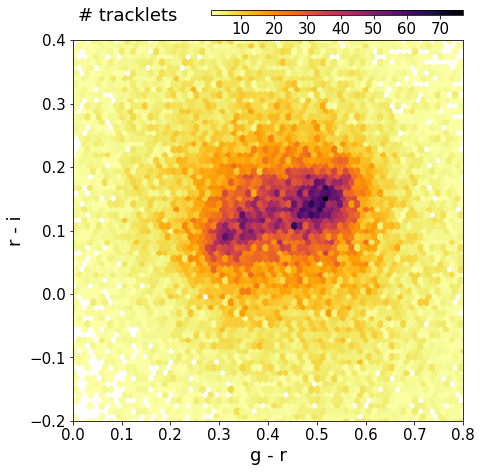}
\caption{Color-color density plot of the 50,222 tracklets found by CANFind with \textit{g}, \textit{r}, and \textit{i} band measurements.  In particular, two distinct groups (bluer carbonaceous, left and redder siliceous, right) are visible. }
\label{fig:ccd}
\end{figure}

\begin{figure*}[t]
\centering
\includegraphics[width=1.0\hsize,angle=0]{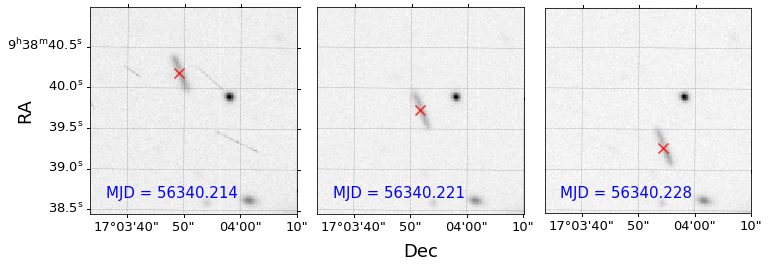}
\caption{\edittwob{An example of a tracklet moving fast enough to leave streaked measurements in each 600-sec. exposure above.  These CP-processed DECam exposures from NSC DR1 show 7\arcsec trails with their Source-Extracted measurements over-plotted in red crosses.  The trail length for the exposure time matches the proper motion predicted by CANFind from the tracklet's trajectory, $\mu$ = 42\arh.   \textit{Note: For those interested, the ``stationary'' object behind the tracklet is white dwarf SDSS J093839.94+170357.7}}}
\label{fig:long_exptime}
\end{figure*}

\item \textit{Analysis of Trailed Detections:}
\edittwob{We plan to identify and analyze streaked detections in NSC exposures to determine further SSO candidates in NSD DR1.  While several tracklets detected by CANFind contain trailed measurements, individual streaks remain unidentified.  These will be identified using new or existing software on exposures and analyzed using PSF fitting.}

\item \textit{Orbit Determination:}
Due to retrograde motion, linking tracks, tracklets, and unclustered measurements to form potential solar system orbits requires the use of an orbit fitting program designed to account for the effects of the Earth's motion on the apparent trajectories of other moving objects.  
Groups of measurements connected by a possible solar system orbit will be designated as a solar system object candidate, and will be studied and reported to the Minor Planet Center.

Properties of derived orbits will be analyzed against known objects.  Many science questions can be addressed, especially considering the great quantity of observations available in the NSC.  
Once all tracks and tracklets in NSC DR1 have been identified using CANFind and the Hough Transform approach, they will be fit to possible SSO orbits.  \editoneb{We hold off on this step until as many tracklets as possible can be identified in the NSC for the purpose of increasing the likelihood of correctly matching tracklets corresponding to unique objects.}

\item \textit{Comparison of Efficiency with THOR:}
The effectiveness of these detection methods will be tested against the results of Tracklet-less Heliocentric Object Recovery \citep[THOR;][]{Moeyens}, which recovers known solar system objects by applying test orbits to object detections.  THOR applies several orbits that are characteristic of distinct types of solar system objects to detections in a region of the sky, and objects placed in very similar orbits are evaluated and connected.  When tested on data from the Zwicky Transient Facility, THOR recovered 97.4\% of known asteroids \citep{Moeyens}.  However, THOR has not yet been tested or optimized for NEOs and cannot generate initial orbits.  We will run THOR on the NSC and compare the resulting set of solar system objects to those detected by the methods described above.
\end{itemize}

\section{Summary}
\label{sec:summary}

We describe CANFind, a new Solar System object detection algorithm which we have applied on DR1 of the NOIRLab Source Catalog.  We detected 524,501 tracklets in 35,644 HEALPix of nside-128, covering 19\% of the sky. 
The majority of tracklets have photometric data in the \textit{g}, \textit{r}, \textit{VR}, and \textit{i} bands, with fewer detections in \textit{u}, \textit{z}, and \textit{y}.  The 50,000 tracklets with measurements in the \textit{g}, \textit{r}, and \textit{i} bands show a bimodality of siliceous and carbonaceous objects, supporting the results of prior studies.  In proper motion space, tracklets fall into groups with motion indicating the presence of Main Belt objects, Jupiter Trojans, Hilda asteroids, and Trans-Neptunian objects.  Roughly one third of all tracklets were detected within 5$^{\circ}$ of the ecliptic.  

In the near future, CANFind will be run on NSC DR2 which contains 68 billion measurements of 3.9 billion unique objects \citep{Nidever2020}.  The data will also be searched using a modified Hough Transform approach in order to link tracklets with detections that are farther spaced-out.  Tracklets found will be fit to potential solar system orbits and submitted to the Minor Planet Center.  

\editoneb{The CANFind source code and a complete catalog of the tracklets detected in the NSC DR1 are publicly available at \url{https://github.com/katiefasbender/CANFind}.  The tracklet catalog includes information about each tracklet's mean (\textit{RA,Dec}) position, date of first observation, average magnitude in each of the NSC's filters, proper motion in both equatorial and ecliptic coordinates [\ded], in FITS format.}

\section{Acknowledgements}
\begin{acknowledgments}
This work is made possible by the National Science Foundation's National Optical-Infrared Research Laboratory (https://nationalastro.org) and the National Optical Astronomy Observatory (https://www.noao.edu).  Computational efforts were performed on the Hyalite High Performance Computing System, operated and supported by University Information Technology Research Cyber infrastructure at Montana State University.

\editoneb{We thank the anonymous reviewer for the detailed and thorough referee report; the overall quality and scientific merit of this manuscript is vastly improved by their recommendations.}

\software{
    \package{Astropy} \citep{astropy},
    \package{IPython} \citep{ipython},
    \package{matplotlib} \citep{mpl},
    \package{numpy} \citep{numpy},
    \package{scipy} \citep{scipy},
    \package{healpy} \citep{Zonca2019},
    \package{scikit-learn} \citep{scikit-learn}
}

\facilities{CTIO:Blanco (DECam), KPNO:Mayall (Mosaic-3), Steward:Bok (90Prime), Astro Data Lab}

\end{acknowledgments}

\bibliographystyle{aasjournals}



\end{document}

%% file: shorthand.tex
\newcommand{\arh}{\arcsec hr$^{-1}$}
\newcommand{\ded}{$^{\circ}$day$^{-1}$}

\definecolor{gold}{rgb}{.8,.6,0}

\newcommand{\package}[1]{\textsl{#1}}

\newcommand{\editone}[1]{\textcolor{red}{\textbf{#1}}} 
\newcommand{\eonesout}[1]{}

\newcommand{\editoneb}[1]{\textcolor{black}{#1}}

\newcommand{\etwosout}[1]{}
\newcommand{\edittwob}[1]{\textcolor{black}{#1}}

\newcommand{\ethree}[1]{\textcolor{black}{#1}}
\newcommand{\ethreesout}[1]{}

%% file: authors.tex
\correspondingauthor{Katie M. Fasbender}
\email{katiefasbender@montana.edu}

\author[0000-0003-1622-5901]{Katie M. Fasbender}
\affiliation{Department of Physics, Montana State University, P.O. Box 173840, Bozeman, MT 59717-3840}

\author[0000-0002-1793-3689]{David L. Nidever}
\affiliation{Department of Physics, Montana State University, P.O. Box 173840, Bozeman, MT 59717-3840}
\affiliation{NSF's National Optical-Infrared Astronomy Research Laboratory, 950 North Cherry Ave, Tucson, AZ 85719}

\shortauthors{Fasbender et al.}